\documentclass{ws-ijmpcs}
\usepackage{braket,psfrag}
\usepackage{bm}

\def\trimmarks{}

\begin{document}

\markboth{I.O. Cherednikov, T. Mertens, F.F. Van der Veken}
{Loop space and evolution of the light-like Wilson polygons}

%
\catchline{}{}{}{}{}
%

\title{LOOP SPACE AND EVOLUTION OF THE LIGHT-LIKE WILSON POLYGONS}

\author{I.O. CHEREDNIKOV\footnote{
On leave of absence from BLTP JINR, Dubna, Russia}}

\address{Departement Fysica, Universiteit Antwerpen, B-2020 Antwerpen, Belgium\\
igor.cherednikov@ua.ac.be}

\author{T. MERTENS}

\address{Departement Fysica, Universiteit Antwerpen, B-2020 Antwerpen, Belgium\\
tom.mertens@ua.ac.be}

\author{F.F. VAN DER VEKEN}

\address{Departement Fysica, Universiteit Antwerpen, B-2020 Antwerpen, Belgium\\
frederik.vanderveken@ua.ac.be}

\maketitle

\begin{history}
\received{\today}
\end{history}

\begin{abstract}
We address a connection between the energy evolution of the polygonal light-like Wilson exponentials and the geometry of the loop space with the gauge invariant Wilson loops of a variety of shapes being the fundamental degrees of freedom. The renormalization properties and the differential area evolution of these Wilson polygons are studied by making use of the universal Schwinger quantum dynamical approach. We discuss the appropriateness of the dynamical differential equations in the loop space to the study of the energy evolution of the collinear and transverse-momentum dependent parton distribution functions.

\end{abstract}


\section{Introduction}
\label{sec:intro}
Transverse-momentum dependent parton distribution functions (TMDs in what follows) are known to possess more involved singularity structure than that of the integrated collinear parton densities\cite{TMD_singular}. Extra divergences which arise in the TMDs due to the light-like Wilson lines (either, equivalently, in the light-cone axial gauge) affect as well the renormalization properties of these objects\cite{WL_LC_rapidity,CS_all} (see also Ref.~\refcite{SCET_TMD} for the discussions of similar issues in some different contexts). In particular, the additional rapidity singularities question the sufficiency of the standard $R-$operation to the renormalization of the TMDs with the light-like Wilson lines.

On the other hand, the cusped Wilson loops defined completely or partially on the light-cone are also known to lack the multiplicative renormalizability due to the arising of the specific light-cone singularities in addition to the common ultraviolet and the infrared ones\cite{WL_LC_rect}. Still a renormalization-group equation can be written down. The cusp anomalous dimension which enters this equation is known to be of remarkable universality\cite{KR87,CAD_universal,St_Kr_Pheno}. It may be helpful, therefore, to study the features and structures which are common to these seemingly different objects. We will concentrate on those which arise from the presence of the light-cone Wilson lines and display themselves in the ``too singular'' pole terms.

\section{Dynamics of the light-like Wilson polygons and area/energy evolution}
To be more precise in the analysis of the singularities and the renormalization properties of matrix elements with the light-like objects, let us consider the Wilson loop defined on a rectangular contour with the sides $N^+, N^-$ along the light-like rays.
Study of the Wilson rectangles on the light-cone is also motivated by the recently observed duality between the $n-$gluon scattering amplitudes in the ${\cal N} = 4 $ super-Yang-Mills theory and the vacuum expectation values of the Wilson loops constructed from $n$ light-like segments connecting the points ${x_i}$, where the lengths of these segments are chosen to be equal to the momenta of the gluon legs $x_i - x_{i+1} = p_i$ (see, e.g.,  Refs.~\refcite{WL_CFT}). The infrared singularities of the ${\cal N} = 4$ amplitudes are supposed to have their counter-terms in the ultraviolet poles of the Wilson polygon. The cusp anomalous dimension\cite{KR87} is the crucial ingredient of the corresponding evolution equations. In other words, the dynamical core of a scattering process in the momentum space reveals itself in the local properties of a polygonal light-like Wilson loop in the coordinate space. Finally, the local features of Minkowskian paths near the obstruction points can be formulated in terms of the universal cusp anomalous dimension.

In particular, in the large-$N_c$ limit we have in the coordinate space\cite{WL_LC_rect}
\begin{eqnarray}
& & W(\Gamma_\Box)  = 1 - \frac{1}{\epsilon^2}\ \frac{\alpha_s N_c}{2\pi} \
\left(\left[\frac{-2 N^+ N^- + i0}{\mu^2}\right]^\epsilon + \left[\frac{2 N^+ N^- + i0}{\mu^2}\right]^\epsilon  \right)  \label{eq:WL_LC_1loop} \\ & &  + \frac{\alpha_s N_c}{2\pi} \left( \frac{1}{2} \ln^2 \frac{N^+N^-}{-N^+N^-} + 2 \zeta_2 + O(\epsilon) \right) + O(\alpha_sN_c) \, . \nonumber
\end{eqnarray}
The function (\ref{eq:WL_LC_1loop}) is dimensionally regularized, and contains a double-pole term of the order $1/\epsilon^2$, which makes it (straightforwardly) non-renormalizable. It is possible, however, to construct a consistent renormalization procedure even for such a ``too singular'' object.

To this end, let us define the area differentials in the transverse $\vec z_\perp = 0$  :
\begin{equation}
{ \delta \sigma^{+-} }
=
{ N^+ \delta N^- } \to p_1 \delta p_2 = \frac{1}{2} \delta s \ ; \
 { \delta \sigma^{-+} }
=
- { N^- \delta N^+ } \to - p_2 \delta p_1 = \frac{1}{2} \delta t \ ,
\label{eq:delta_area}
\end{equation}
where momentum variables $s,t$ remind us about the duality between the coordinates $N^\pm$ and momenta of the corresponding external gluons in the $N=4$ SYM scattering amplitude. It is worth noting that these operations are defined strictly in the corners ${x_i}$, and we distinguish between ``left'' and ``right'' variations, see Fig. 1.

\begin{figure}[ht]
 $$\includegraphics[angle=90,width=0.7\textwidth]{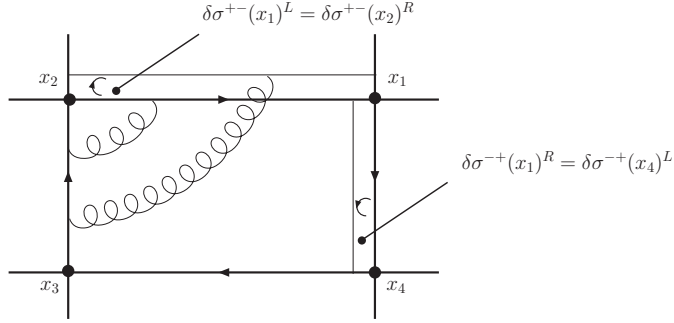}$$
   \caption{Area variations allowed for a light-cone Wilson rectangle: we consider only those area variations which conserve the angles between the sides.}
\end{figure}
The renormalizability is jeopardized because of the higher ``degree of singularity'' caused by the light-like Wilson lines.
In order to decrease the degree of singularity, we make use of the approach described in Refs.~\refcite{CAD_universal}. Applying the
area logarithmic derivation operator
\begin{eqnarray}
  \frac{\delta}{\delta \ln \sigma}
  \equiv
  \sigma_{\mu\nu} \frac{\delta}{\delta \sigma_{\mu\nu}}
  =
  \sigma_{+-} \frac{\delta}{\delta \sigma_{+-}}
  +
  \sigma_{-+} \frac{\delta}{\delta \sigma_{-+}} \
  \label{eq:area_log}
\end{eqnarray}
to the r.h.s. of the (\ref{eq:WL_LC_1loop}), we obtain
\begin{eqnarray}
& & \frac{\delta}{\delta \ln \sigma_{\mu\nu}} \ \ln W(\Gamma_\Box)
= - { \frac{\alpha_s N_c}{2\pi} } \ \frac{1}{\epsilon}   \left( \left[\frac{-2N^+N^-  + i0}{\mu^2}\right]^\epsilon + \left[\frac{2N^+N^- + i0}{\mu^2}\right]^\epsilon \right) .
\label{eq:log_der}
\end{eqnarray}
The {cusp anomalous dimension} arises now after additional logarithmic differentiation in the dimensional regularization scale $\mu$
\begin{equation}
 \mu \frac{d}{d\mu} \frac{\delta \ \ln W(\Gamma_\Box)}{\delta \ln \sigma}
 =
 - 4 \ { \Gamma_{\rm cusp} } \ , \ {\Gamma_{\rm cusp} = \frac{\alpha_s N_c}{2 \pi} } \ .
 \label{eq:full_der}
\end{equation}
The result in (\ref{eq:full_der}) describes the dynamical properties of the light-like Wilson rectangle in terms of the differential area (cusp angles conserving) transformations, allowing us to relate the {geometry} of the loop space to the {dynamics} of the fundamental degrees of freedom---the gauge invariant, regularization independent {light-like WLs}\cite{ChMVdV_2012}.
Note that  (\ref{eq:full_der}) also agrees with the non-Abelian exponentiation of the regularized Wilson loops:
\begin{equation}
 W (\Gamma_\Box; \epsilon)
 =
 \exp \left[ \sum_{k=1} \alpha_s^k \ C_k (W) F_k (W) \ \right] \ ,
\end{equation}
where the ``maximally non-Abelian'' numerical coefficients are $
 {C_k} \sim C_F \ N_c^{k-1} \to {\frac{N_c^k}{2} } \ , $
and the summation goes over all ``webs'' $F_k$, see  Refs.~\refcite{WL_expo,KR87}.
The above observation is not surprising: it follows directly from the property of linearity of the cusp anomalous dimension in the large-angle limit\cite{KR87}. This asymptotical regime is realized exactly in the light-cone case.

It is instructive to consider another example: the $\Pi-$shape Wilson (semi-)loop with one of the segments lying on the light-cone and two semi-infinite off-light-cone sides, Fig. 2.
\begin{figure}[ht]
 $$\includegraphics[angle=90,scale=0.5]{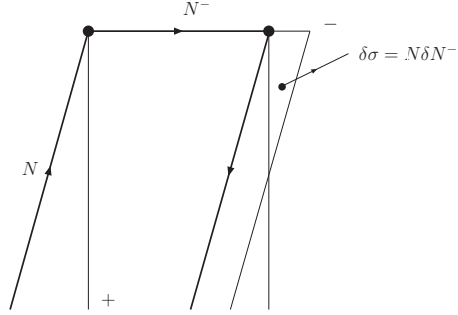}$$
   \vspace{0.2cm}
   \caption{\label{fig:2}$\Pi$-shape Wilson contour and the infinitesimal area variations.}
\end{figure}
In the 't Hooft limit, we have\cite{WL_Pi}
\begin{eqnarray}
  & & W(\Gamma_\Pi)
  =
  1 + \frac{\alpha_s N_c}{2\pi} \  +
  \left[ - L^2 (NN^-) + L (NN^-)  - \frac{5 \pi^2}{24}  \right] \ , \nonumber \\
  & & L(NN^-)
  = \frac{1}{2}\left(\ln (\mu N N^- + i0) + \ln (\mu NN^- + i0) \right)^2 \ ,
  \label{eq:pi_1loop}
  \end{eqnarray}
where the area is given by the product of the light-like $N^-$ and non-light-like $N$ vectors, and area differentials are defined in Fig. 2.
The $\Pi$-shaped Wilson loop (\ref{eq:pi_1loop}) obeys Eq. (\ref{eq:mod_schwinger}), Refs.~\refcite{ChMVdV_2012}:
\begin{equation}
  \mu \frac{d }{d \mu } \ \left[ \frac{d}{d \ln \sigma} \ \ln \ { W(\Gamma_\Pi) } \right]
  =
  - 2 { \Gamma_{\rm cusp} } \ .
  \label{Eq:pi_der}
\end{equation}
The cusp anomalous dimension controls, therefore, the renormalization properties of the integrated parton densities at large-$x$ and the anomalous dimensions of conformal operators with large Lorentz spin\cite{WL_Pi}.
It is remarkable that the TMDs with the longitudinal gauge links on the light-cone $\Phi (x, \vec k_\perp)$ display the similar behavior
\begin{equation}
  \mu \frac{d }{d \mu } \ \left[ \frac{d}{d \ln \theta} \ \ln \ { \Phi (x, \vec k_\perp) } \right]
  =
  2 { \Gamma_{\rm cusp} } \ ,
  \label{eq:tmd_der}
\end{equation}
where the information about the area transformations is accumulated in the rapidity cutoff $\theta = \eta/p^+$\cite{CS_all}.

Indeed, the energy/area evolution equations follow from the fundamental {quantum dynamical principle} by Schwinger\cite{Schwinger51}. According to the Schwinger approach, the quantum action operator $S$ controls variations of arbitrary states
\begin{equation}
 {  \delta \langle \ \alpha' \ |\ \alpha''\  \rangle }
 =
 \frac{i}{\hbar} {  \langle \ \alpha' \ | \delta S | \ \alpha'' \ \rangle } \ .
 \label{eq:principle}
\end{equation}
However, Eq. (\ref{eq:principle}) knows nothing about singularities of the objects to which it is supposed to be applicable. In particular, direct use of the Schwinger methods (supplied with the Stokes theorem) yields the Makeenko-Migdal loop equations, missing the information about light-cone cusp singularities. Let us try to avoid the operations which suggest the smoothness of the Wilson loops under consideration.
The results obtained in the previous Section imply that the study of the area variations (\ref{eq:delta_area}) may be of some use
\begin{equation}
 {  \frac{\delta}{\delta \sigma} \langle \ \alpha' \ |\ \alpha''\  \rangle }
 =
 \frac{i}{\hbar} {  \langle \ \alpha' \ | \frac{\delta}{\delta \sigma} S | \ \alpha'' \ \rangle } \ .
 \label{eq:principle_mod}
\end{equation}
Taking into account the renormalization group invariance of the Schwinger formmula, we obtain for the cusped light-like contours:
\begin{equation}
 \mu \frac{d }{d \mu } \ {\left[ \sigma_{\mu\nu} \frac{\delta}{\delta \sigma_{\mu\nu}} \ \ln \ W (\Gamma) \right] }
 =
 - \sum { \Gamma_{\rm cusp} } \ .
 \label{eq:mod_schwinger}
\end{equation}
Thus, Eqs. (\ref{eq:full_der}, \ref{Eq:pi_der}, \ref{eq:tmd_der}) are some of the particular examples of the generalized Schwinger approach, Eq. (\ref{eq:mod_schwinger}).
Note, however, that the r.h.s. of the Eq. (\ref{eq:mod_schwinger}) is not derived, strictly speaking, from the first principles: we have used the explicit results for the known quantities and continued them to arbitrary elements of the loop space. More detailed discussion of the derivation will be reported elsewhere.

\section{Outlook}
We have studied some properties of the Wilson light-like polygons, the latter being
the elements of the generic loops space and, correspondingly, the fundamental degrees of freedom of the gauge-invariant formulation of QCD\cite{Loop_Space,WL_RG,St_Kr_WL_cast}:
\begin{eqnarray}
& &  W_n (\Gamma_1, ... \Gamma_n)
  =
\Big \langle 0 \Big| {\cal T} \frac{1}{N_c} {\rm Tr}\ \Phi (\Gamma_1)\cdot \cdot \cdot \frac{1}{N_c}{\rm Tr}\ \Phi (\Gamma_n)  \Big| 0 \Big\rangle \ , \\
& &  \Phi (\Gamma_i)
   =
   {\cal P} \ \exp \left[ ig \oint_{\Gamma_i} \ dz^\mu A_{\mu} (z) \right] \ \nonumber
   \label{eq:wl_def} \ .
\end{eqnarray}
It is known since the late 70's that the general dynamics of these objects is described by the Makeenko-Migdal (MM) equations\cite{MM_WL,WL_Renorm}, which can be obtained by applying the Schwinger-Dyson approach to the scalar functionals $\Phi (\Gamma)$  (\ref{eq:wl_def}):
\begin{equation}
 \partial_x^\nu \ \frac{\delta}{\delta \sigma_{\mu\nu} (x)} \ W_1(\Gamma)
 =
 N_c g^2 \ \oint_{\Gamma} \ dz^\mu \ \delta^{(4)} (x - z) W_2(\Gamma_{xz} \Gamma_{zx}) \ ,
 \label{eq:MM_general}
\end{equation}
where the definitions of the path and area differential operations are given, e.g., in the Refs.~\refcite{MM_WL,WL_Renorm}.
However, the Eqs. (\ref{eq:MM_general}) are of somewhat limited practical use due to, in particular, extra divergences\cite{WL_Renorm}, emerging from a variety of the obstructions, or from the light-like segments of the integration contours.
There are also subtleties related to the continuous deformations of paths in Minkowski space-time, which is argued in the Refs.~\refcite{Topo_ST} to be as unconnected as a space can be with respect to a path-topology, making the meaning of the derivatives unclear.

Handling these problems will require a deep understanding of what causes them. The different objects in the Eq. (\ref{eq:MM_general}) hint that the divergence and derivative definition problems either originate from the contours themselves or from the gauge fields defined on these contours. Consider for a moment a rectangular closed contour. Topologically, this contour is identical to $S^1$, thus although the original contour contains cusps they do not form any obstructions from a purely topological point of view. The problems with the cusps arise when considering derivatives, what means that one has assumed extra structure on the contours such as embedding them in another space (Minkowski, Euclidean), a parametrization, a metric/gauge field.  From the physical point of view, something happened at the cusp, a particle gets scattered, implying that we shall need to introduce something (like a vertex function or an operator) to take this into account when studying the contractions of the gauge fields belonging to opposite sides of our rectangular example.

Inspired by the example we split up our investigation of the MM equations into the study of the topological classes of the contours and of the extra structure generated by the gauge fields, the embeddings, etc. The topology of the contours has lead us to the knot theory, which is able to provide the consistent treatments of the different classes of contours and their properties (group operations, linking, etc.). The study of the extra structure implies the use of the concept of the generated (emergent) space-times and finite-universe topologies. In particular, in the twistor theory, the complexified and compactified space-time is considered to be generated from the twistor space. The advantage of the twistor theory is that the differential data are exchanged for the algebraic data. Moreover, recently a twistor theory version of the MM equations (\ref{eq:MM_general}) have been derived introducing the concept of the holomorphic linking, leading to a complexified knot theory, motivating a further study of the twistor and knot theories:
\begin{eqnarray}
 & & \overline{\delta} \left< W \left[ C(t)  \right]\right> =
-\lambda \int\limits_{C(t) \times C(t)} \omega(z) \wedge \omega(z') \wedge \overline{\delta}^{3 \vert 4}(z,z') \left< W \left[C' \left( t \right) \right]\right> \left< W \left[C''\left( t \right) \right]\right> \label{eq:WL_TT} \\
& & - \lambda \int\limits_{\Gamma \times S^1 \times S^1}D^{3 \vert 4}z_a \wedge D^{3 \vert 4}z_b
\left[ \int\limits_{C(t) \times X} \omega(z) \wedge \omega(\hat{z}) \wedge \overline{\delta}^{3 \vert 4}(z,\hat{z})
\left< W\left[\widetilde{C(t)} \cup X \right]\right>\right] \nonumber \ ,
\end{eqnarray}
for detailed explanation and description of the notations, see Refs.~\refcite{WL_TT_ScA_2011}.
Problems with the behavior of the Wilson lines/loops at infinity drive us to consider alternatives to the usually applied one-point compactification of the space-time. For instance, the sign difference caused by the oppositely directed Wilson lines in the TMDs for the Drell-Yan and semi-inclusive Deep Inelastic Scattering processes might suggest that,  from the point of view of the compactification, the point at $-\infty$ is not the same as the point at $+\infty$. Given the above motivation, it seems attractive to look into the finite-universe topologies, which have recently been investigated in the light of CMB data\cite{Finite_Universe}.

In the present work we followed another, more simple, strategy. We made use of the observation that
in the large-$N_c$ limit, in the null-plane, for the light-like {dimensionally regularized (not renormalized)} Wilson rectangles, the area differentials can be reduced to the normal ones. The area differential equations (which can be treated as the non-renormalizable relatives of the MM Eqs.) in the coordinate representation show up the duality to the {energy (or rapidity, e.g., in the TMD case) evolution equations} for the light-like Wilson polygons in the momentum picture. As the result, the obtained differential energy/area equations form a closed set of the dynamical equations for the loop functionals, they can be, in principle, formulated consistently and even solved on the light-cone.
Hence we made some progress in understanding of the relationship between the geometrical properties of the loop space in terms of the area differential evolution equations,, from one side, and the dynamics accumulated in the cusps---the angles between the light-like straight lines, from another side. Thus, in the loop space, the (external) dynamics can be taken into account by introducing the obstructions to the initially smooth loops, with those obstructions resembling the {\it sources} within the Schwinger field-theoretical picture. We have demonstrated that the universal Schwinger quantum dynamical principle is a useful tool to study some special classes of the elements of the loop space, in particular, the cusped Wilson exponentials (null-polygons) on the light-cone. More involved cases, e.g., the non-light-like Wilson polygons and polyhedra deserve separate study\cite{WL_Polygons}.

\paragraph{Acknowledgements}
We thank the participants of the JLab QCD Evolution workshop and of the summer meetings in ECT*, Trento, for useful discussions. We appreciate valuable comments and critical remarks by I.V. Anikin, Y.M. Makeenko and N.G. Stefanis.



\end{document}